%% file: main.tex
\documentclass[conference]{IEEEtran}
\IEEEoverridecommandlockouts
\usepackage{cite}
\usepackage{amsmath,amssymb,amsfonts}
\usepackage{algorithmic}
\usepackage{graphicx}
\usepackage{textcomp}
\usepackage{xcolor}

\usepackage{float}
\usepackage{multirow}
\usepackage{hyperref}
\usepackage{soul}
\usepackage{array}
\newcolumntype{L}[1]{>{\raggedright\let\newline\\\arraybackslash\hspace{0pt}}p{#1}}

\def\BibTeX{{\rm B\kern-.05em{\sc i\kern-.025em b}\kern-.08em
    T\kern-.1667em\lower.7ex\hbox{E}\kern-.125emX}}
\begin{document}

\title{Define-ML: An Approach to Ideate Machine Learning-Enabled Systems}

\author{\IEEEauthorblockN{Silvio Alonso, Antonio Pedro Santos Alves, Lucas Romao, Hélio Lopes, Marcos Kalinowski}
\IEEEauthorblockA{\textit{Department of Informatics} \\
\textit{Pontifical Catholic University of Rio de Janeiro (PUC-Rio)}\\
Rio de Janeiro, Brazil \\
\{smarques, lopes, kalinowski\}@inf.puc-rio.br}

}

\maketitle

\input{abstract}


\input{1_introduction}
\input{2_background}

\input{XXX_framework_proposal}
\input{3_design}

\input{4_candidate_solution}
\input{5_static_validation}
\input{6_dynamic_validation}
\input{7_discussion}
\input{8_threats}
\input{9_conclusion}

\section*{Acknowledgment}

We express our gratitude to CNPq (Grant 312275/2023-4), FAPERJ (Grant E-26/204.256/2024), Kunumi, and Stone Co. for their generous support.

\bibliographystyle{./bibliography/IEEEtran}
\bibliography{./bibliography/IEEEabrv,./bibliography/IEEEexample, bibliography/references}

\vspace{12pt}
\end{document}

%% file: abstract.tex
\begin{abstract}
   
    \textbf{[Context] The increasing adoption of machine learning (ML) in software systems demands specialized ideation approaches that address ML-specific challenges, including data dependencies, technical feasibility, and alignment between business objectives and probabilistic system behavior. Traditional ideation methods like Lean Inception lack structured support for these ML considerations, which can result in misaligned product visions and unrealistic expectations. [Goal] This paper presents Define-ML, a framework that extends Lean Inception with tailored activities—Data Source Mapping, Feature-to-Data Source Mapping, and ML Mapping—to systematically integrate data and technical constraints into early-stage ML product ideation. [Method] We developed and validated Define-ML following the Technology Transfer Model, conducting both static validation (with a toy problem) and dynamic validation (in a real-world industrial case study). The analysis combined quantitative surveys with qualitative feedback, assessing utility, ease of use, and intent of adoption. [Results] Participants found Define-ML effective for clarifying data concerns, aligning ML capabilities with business goals, and fostering cross-functional collaboration. The approach's structured activities reduced ideation ambiguity, though some noted a learning curve for ML-specific components, which can be mitigated by expert facilitation. All participants expressed the intention to adopt Define-ML. [Conclusion] Define-ML provides an openly available, validated approach for ML product ideation, building on Lean Inception's agility while aligning features with available data and increasing awareness of technical feasibility.} 

\end{abstract}

%% file: 1_introduction.tex
\section{Introduction}  

As machine learning (ML) continues to permeate software products across diverse sectors, pressure increases to ideate ML-enabled systems that are both innovative and feasible. However, early-stage ideation for such systems is uniquely challenging. ML features are inherently probabilistic, dependent on data quality and availability, and often misunderstood by stakeholders unfamiliar with the limitations of ML~\cite{nahar2023meta}. Without proper guidance, teams risk proposing solutions that are technically infeasible, misaligned with business objectives, or disconnected from the realities of the available data.

Despite the popularity of structured ideation approaches like Lean Inception~\cite{caroli2018lean}, these methods were designed for traditional software product ideation and lack explicit support for ML-specific considerations. They do not offer explicit mechanisms to evaluate data readiness or to align feature ideas with ML capabilities. This is particularly important considering that managing customer expectations and aligning requirements with data are among the main pain points of engineering ML-enabled systems~\cite{alves2023status}\cite{kalinowski2024naming}.

To address this gap, we introduce Define-ML, a framework that extends Lean Inception with three structured activities: Data Source Mapping, Feature-to-Data Source Mapping, and ML Mapping. These activities guide teams in grounding their ideation process in data realities, clarifying the technical feasibility of ML features, and fostering alignment between business stakeholders, domain experts, and ML practitioners.

Define-ML was developed and validated following the Technology Transfer Model~\cite{gorschek2006model}. We conducted a static validation using a simulated problem with industry practitioners and a dynamic validation through a real-world case study in the retail domain. In both settings, participants reported increased clarity, improved alignment, and strong intent to adopt the framework. Define-ML offers a structured approach to ideate ML-enabled systems, helping to connect business vision and ML feasibility from the beginning of product development.

The remainder of this paper is organized as follows. Section~\ref{sec:background} presents the background and related work. Section~\ref{sec:methodology} presents a description of the research methodology. Sections~\ref{sec:candidate_solution} presents the Define-ML approach.
Sections~\ref{sec:static_validation} and~\ref{sec:dynamic_validation} describe the static and dynamic validations. Section~\ref{sec:discussion} discusses the results and threats to validity. Finally, Section~\ref{sec:conclusion} concludes the paper.

%% file: 2_background.tex
\section{Background and Related Work}\label{sec:background}

Several ideation approaches exist for software products, \textit{e.g.}, Design Thinking~\cite{brown2008design}, Design Sprints~\cite{banfield2015design}, Lean Startup~\cite{ries2011lean}, and Lean Inception~\cite{caroli2018lean}. We adopt Lean Inception as our foundation due to its structured yet flexible approach to product visioning and feature prioritization, qualities particularly valuable for ML-enabled systems, where technical feasibility must align with business goals. In this section, we provide the background on the Lean Inception workshop and review related work on ideating ML-enabled systems.

\subsection{Lean Inception}

Lean Inception is an ideation workshop that fosters teamwork and collaboration~\cite{caroli2018lean}. It involves assembling a cross-functional team, encompassing developers, designers, business analysts, and representatives of different stakeholders. The process is conducted in a workshop format, usually spanning about a week, to ensure a concentrated and intensive period of discussion, brainstorming, and decision-making.

Creating a shared product vision is a significant outcome of the Lean Inception process. This vision becomes the guiding beacon for all subsequent activities and decisions. A central element of the approach is the deep understanding and prioritization of customer needs. This is achieved by identifying user personas and delving into their problems and challenges. This understanding is important to ensure that the project or product directly addresses user requirements. Alongside this, journey mapping is employed to visualize and understand the user’s experience, identifying key features and interaction points. 

An important result of a Lean Inception workshop is the prioritization of product features. The team works to identify the most critical elements for the initial product version, often termed the Minimum Viable Product (MVP), and decides what can be developed later.

\subsection{Ideation of ML-enabled products}
Jansen and Colombo~\cite{jansen2023mix} address the challenges designers face in integrating ML into their processes. They propose the \textit{Mix \& Match Machine Learning Toolkit}, which aims to support ML-enabled ideation by providing tangible and accessible ML knowledge. The toolkit consists of physical tokens that represent data types (\textit{e.g.}, labeled/unlabeled audio, images, text) and ML capabilities (\textit{e.g.}, categorize, recommend, generate), along with a web interface for exploring and combining elements. Leveraging Tangible User Interfaces (TUIs), the approach fosters exploration and collaboration among designers, making ML concepts more approachable without requiring deep ML expertise. Workshops with design students showed that the toolkit supported both learning and ideation. It helped participants grasp basic ML concepts and generate novel ideas by offering a structured approach. It also influenced their language and thinking, as they adopted ML-related terminology during ideation.

Yildirim \textit{et al.}~\cite{yildirim2023creating} address AI integration challenges through human-centered design, developing a resource catalog of 40 common AI features across domains. Their study revealed that moderate AI performance often delivers more practical value than high-performance solutions. The authors created a flexible design tool featuring examples and an AI capability grammar, which effectively expanded idea generation in design sessions. A complementary study~\cite{yildirim2023investigating} examined how practitioners use human-AI guidelines (more specifically, the \textit{People+AI Guidebook}) when designing AI-enabled products. The findings emphasize the need for enhanced problem-framing tools to support initial design phases. Although the guidelines are valuable for education and communication, practitioners explicitly reported insufficient early-stage ideation support. 

%% file: 3_design.tex
\section{Research Methodology}\label{sec:methodology}

\begin{figure}[ht]
    \centering
    \includegraphics[width=0.35\textwidth]{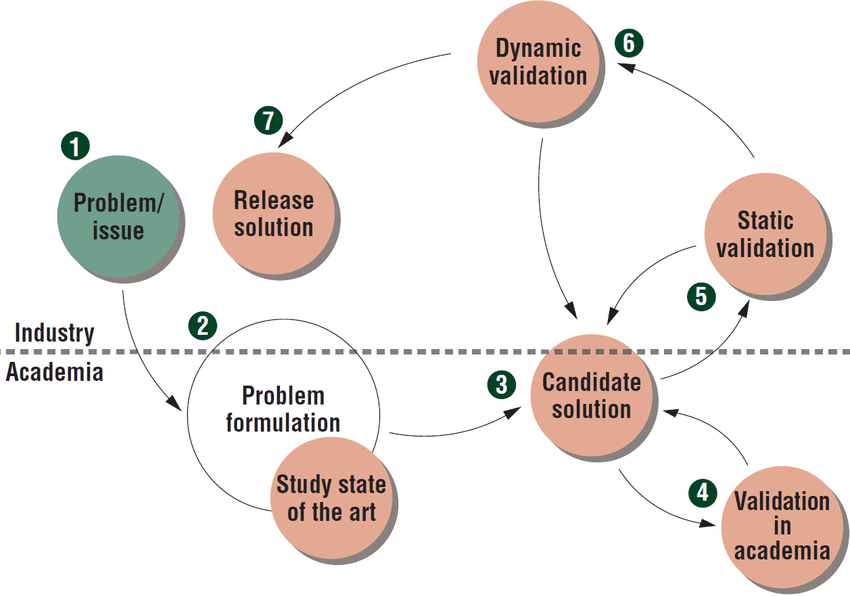}
    \caption{Technology transfer model proposed by Gorschek \textit{et al.}~\cite{gorschek2006model}}
    \label{fig:techonology_transfer_model}
\end{figure}

We followed the seven steps proposed by the Technology Transfer Model~\cite{gorschek2006model} and describe them hereafter. Figure~\ref{fig:techonology_transfer_model} presents how the steps interact with each other. The instruments and data collected in each conducted study are available in our open science repository\footnote{\url{https://doi.org/10.5281/zenodo.15277758}}.

\subsection{Step 1. Identify problem based on industry needs}
The ideation and development of ML-enabled products pose unique challenges distinct from traditional software systems. A key difficulty is aligning requirements with available data, complicated by ML’s non-deterministic nature~\cite{alves2023status}. Managing stakeholder expectations is also challenging, often due to limited AI literacy~\cite{nahar2023meta}. Communicating ML limitations and uncertainties~\cite{alves2023status}, along with translating vague business goals into concrete functional requirements~\cite{nahar2023meta}\cite{villamizar2024identifying}, further complicates the process.

Several MVP ideation practices, such as Lean Inception and Design Thinking, have been cataloged in a recent systematic mapping~\cite{alonso2023systematic}. However, these approaches lack explicit support for ML-specific concerns—such as data dependencies, model feasibility, and the need to align ML capabilities with business goals and user needs~\cite{ferreira2023lessons}. While some recent efforts address ML component specification~\cite{jansen2023mix}\cite{  yildirim2023creating}, they do not provide a comprehensive framework tailored to the ideation of ML-enabled systems. Our R\&D experiences developing ML-enabled systems with industry partners~\cite{kalinowski2020lean}\cite{KalinowskiRRBVBL25} reinforce these observations of the literature from a practical perspective. We also found that traditional ideation approaches fall short when applied to ML-enabled systems, underscoring the need for adapted methods that explicitly address ML integration while preserving agility and user-centricity.

\subsection{Step 2. Formulate the research problem}

Based on the issue described in the previous step, we defined our research problem as introducing and evaluating Define-ML, an approach that extends Lean Inception to support ML-enabled system ideation. Define ML includes tailored activities like Data Source Mapping, Feature-to-Data Source Mapping, and ML Mapping, helping teams identify necessary data sources and suitable ML capabilities, ensuring that the ideated ML capabilities are not only aligned with business goals but also technically feasible.

\subsection{Step 3. Formulate a candidate solution}

To better support the ideation of ML-enabled products, we propose an adaptation of the Lean Inception approach. The main change consists of introducing three new activities: \textit{Data Source Mapping}, which aims at identifying and discussing the quality of available data sources; \textit{Feature-to-Data Source Mapping}, which connects the features to the necessary data sources to implement them; and \textit{ML Mapping}, an adaptation of the Mix and Match ML toolkit. The latter guides teams in selecting appropriate ML techniques by mapping problem requirements to feasible algorithmic solutions. These extensions address gaps in traditional Lean Inception workshops, which often overlook technical constraints and data dependencies critical to ML projects. The candidate solution is explained in detail in Section~\ref{sec:candidate_solution}.

\subsection{Step 4. Conduct lab validation}

As an initial academic validation, a two-hour workshop was conducted to assess the Define-ML approach. The session took place at the ExACTa PUC-Rio lab and was attended by 18 participants (three professors, six Ph.D. students, six MS students, and three of the lab's hired professionals engaged in industry-academia collaborations). This group brought a wealth of experience, particularly in ideation workshops such as Lean Inception, as many participants had previously contributed to projects using the Lean R\&D framework~\cite{kalinowski2020lean}.

During the workshop, the first author gave a detailed presentation of the Define-ML approach, with a specific focus on the new proposed activities. The presentation aimed to provide a comprehensive understanding of the approach while inviting critical feedback from attendees. Participants were encouraged to share their perspectives on the strengths, limitations, and potential areas for refinement of the approach. The participants agreed with the (theoretical) suitability of the approach and offered constructive suggestions, which resulted in improvements in the artifacts before the studies involving industry partners. For example, the data source mapping artifact was refined to consider perceived quality.

\subsection{Step 5. Perform static validation}

The static validation consisted of a three-hour session at an industry partner, a prominent company in the Brazilian energy sector, in which they were introduced to the Define-ML approach and applied its new activities in practice to a toy problem. The goal was to validate the Define-ML approach with industry practitioners. The static validation and its results are explained in detail in Section~\ref{sec:static_validation}.

\subsection{Step 6. Perform dynamic validation}
The dynamic (real-world) validation of our proposed approach involved conducting a three-day workshop facilitated by the research team and attended by practitioners (data intelligence team) from a multinational energy drink company. The demand from the business partner involved facilitating the ideation of a solution for retail demand forecasting. The workshop strictly followed the Define-ML approach, allowing for the gathering of additional insights on the practical applicability in real-world settings. We document this case study following Runeson \textit{et al.}'s guidelines~\cite{runeson2012case} and present it in detail in Section~\ref{sec:dynamic_validation}.

\subsection{Step 7. Release the solution}

To facilitate the adoption of the Define-ML approach, the researchers released an open-access template in Miro\footnote{\url{https://miro.com/miroverse/defineml-template/}} accompanied by detailed usage instructions. This template is supplemented with an online example demonstrating a common ML-enabled product use-case.

%% file: 4_candidate_solution.tex
\section{Candidate Solution}\label{sec:candidate_solution}

\begin{figure}[ht]
    \centering
    \includegraphics[width=0.45\textwidth]{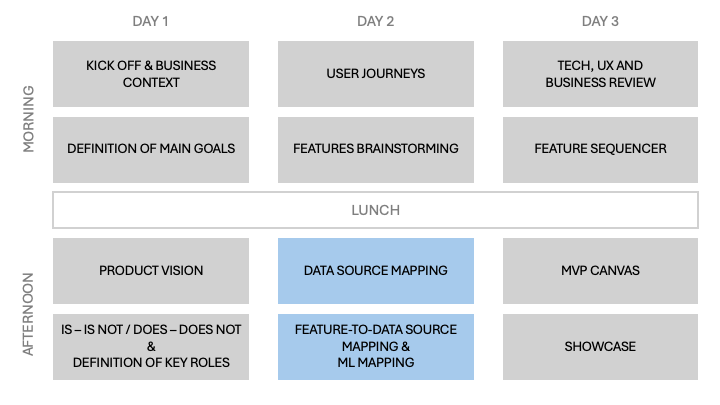}
    \caption{Define-ML typical agenda}
    \label{fig:agenda}
\end{figure}

Define-ML is an adaptation of the Lean Inception approach, that despite providing a solid foundation for cross-functional alignment, lacks activities tailored to the data-specific requirements of ML products. The proposed approach introduces several adaptations to the Lean Inception methodology to better suit the needs of ML-enabled product ideation. A typical agenda for the Define-ML workshop is presented in Figure~\ref{fig:agenda}. Recognizing that ML products require specialized knowledge and consideration of technical constraints, the approach includes the attendance of ML experts in the workshop sessions.These experts provide quick, valuable insights on technical feasibility and offer explanations about the limitations of ML, ensuring that ideas remain grounded in what is realistically achievable. 
Their presence helps the team make informed decisions early on, preventing potential misalignment between the product vision and ML capabilities.

Another adaptation involves replacing the traditional personas definition activity with a streamlined activity focused on identifying the key roles affected by the product. This simplification aims at saving time in the workshop without sacrificing essential context, allowing participants to focus on defining roles rather than detailed user characteristics.

The main adaptation of Define-ML concerns incorporating three new activities specifically designed to address the unique data and technical requirements of ML products: \textit{Data Source Mapping}, \textit{Feature-to-Data Source Mapping}, and \textit{ML Mapping}. These activities are explained hereafter.

\subsection{Data Source Mapping}
\begin{figure}[H]
    \centering
    \includegraphics[width=0.35\textwidth]{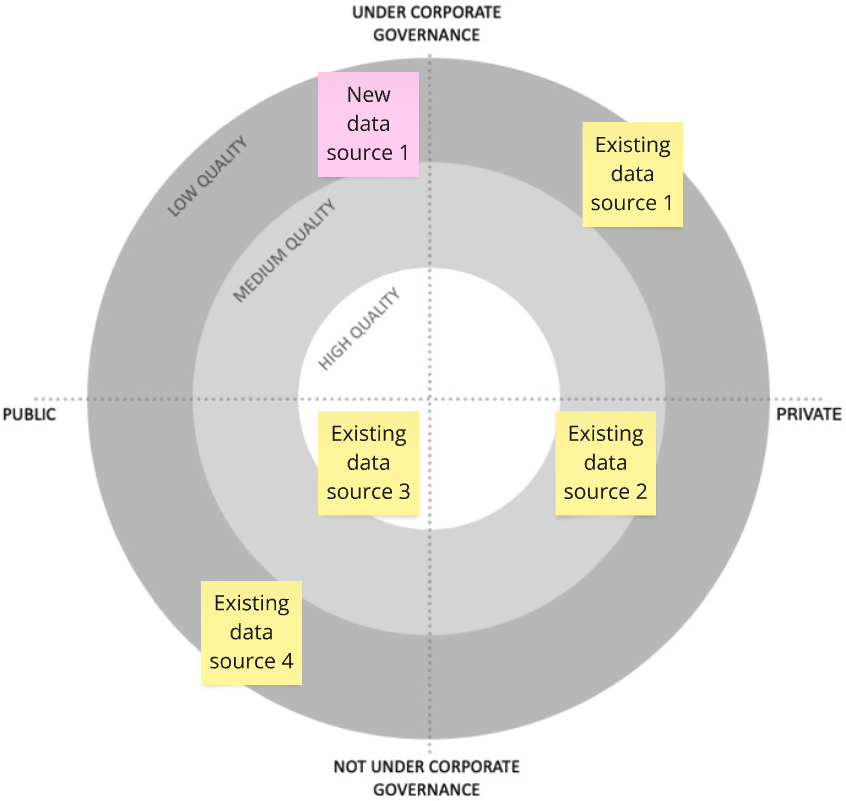}
    \caption{Data source mapping board}
    \label{fig:data_source_mapping}
\end{figure}

The \textit{Data Source Mapping} activity was designed to map the primary data sources used by the organization and the desired data sources relevant for the product. The board used for this mapping can be seen in Figure~\ref{fig:data_source_mapping}.

The board is structured with two axes. The x-axis differentiates whether the data is public or private, helping the team quickly identify access restrictions and privacy considerations. The y-axis distinguishes whether the data is under corporate governance. For instance, data from an ERP system is managed under corporate governance, while a spreadsheet maintained by a specific individual within the organization may not be on corporate governance's radar. Data quality is also visually represented through circles representing three quality levels: high, medium, and low.

Additionally, the approach considers using differently colored post-its to differentiate between data sources the organization already possesses and data sources it wishes to organize or obtain. This allows participants to easily recognize gaps in data availability and identify strategic priorities for acquiring additional data sources necessary for the product. 

\subsection{Feature-to-Data Source Mapping}

\begin{figure}[ht]
    \centering
    \includegraphics[width=0.47\textwidth]{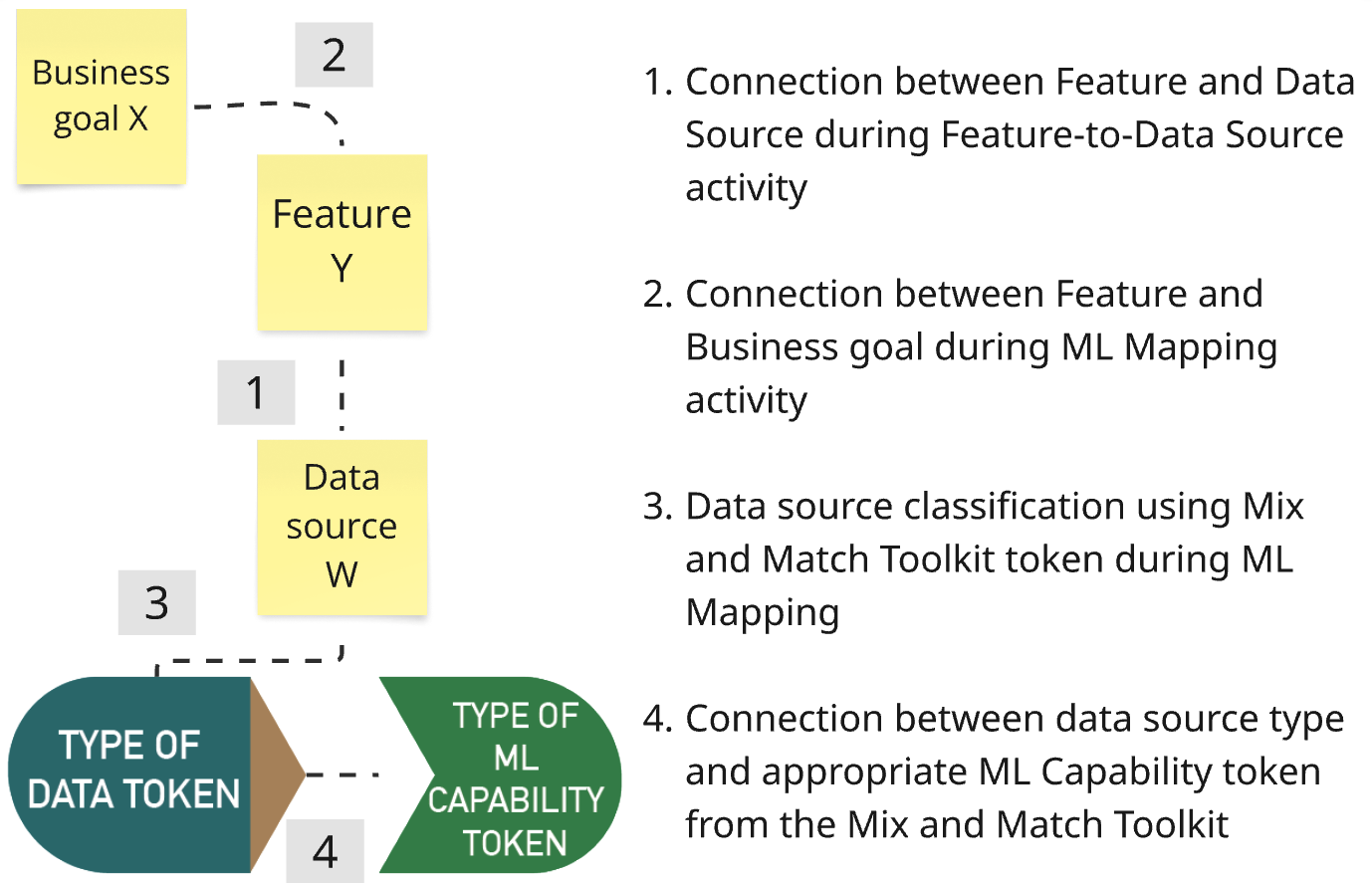}
    \caption{Feature-to-Data Source Mapping and ML Mapping}
    \label{fig:ml_mapping}
\end{figure}

This activity aims to connect the features to the data sources deemed necessary for their development. This step serves to provide a foundational understanding of how data will support the feature’s implementation. Step 1 presented in Figure~\ref{fig:ml_mapping} shows an example of this connection.

\subsection{ML Mapping}

The \textit{ML Mapping} activity starts with categorizing features into ``ML-intensive" (\textit{i.e.}, features that use ML) and ``non-ML-intensive" features. Thereafter, a set of steps is performed for the ML-intensive ones. These steps were inspired by the \textit{Mix \& Match ML toolkit} proposed by Jansen and Colombo~\cite{jansen2023mix}, which involves relating data types to ML capabilities as a way of making complex ML concepts more approachable.

Each ML-intense feature is linked to a business objective. This connection, presented as Step 2 in Figure~\ref{fig:ml_mapping}, addresses the common issue of misalignment between ML initiatives and business objectives, as highlighted by Nahar \textit{et al.}~\cite{nahar2023meta}, ensuring that each ML-intensive feature is purpose-driven and tied to measurable business outcomes.

Following this, the tokens proposed in the Mix and Match ML toolkit~\cite{jansen2023mix} are used, classifying the data sources according to the categories in the toolkit and matching them with appropriate model types. Steps 3 and 4 of Figure~\ref{fig:ml_mapping} represent this classification and matching, respectively. This visual and interactive approach is designed to help participants grasp the technical possibilities, making the gaps between the available data and the desired ML capabilities explicit.

The ML Mapping activity provides a structured approach to align ML capabilities with business objectives, data availability, and ML model possibilities. The involvement of ML experts in this activity is essential, as they help to clarify what could realistically be achieved based on the intended business objectives and on the available data. 

%% file: 5_static_validation.tex
\section{Static Validation}\label{sec:static_validation}

\subsection{Goal and Research Questions}

We defined the goal of the static validation using the Goal-Question-Metric (GQM) goal definition template~\cite{basili1988tame} as follows: ``\textit{Analyze} Define-ML \textit{for the purpose of} characterization \textit{with respect to} perceived usefulness of the new activities and of the approach \textit{from the point of view of} practitioners \textit{in the context of} applying the approach to a toy problem."

Based on this goal, we defined the following Research Questions (RQs): \textbf{RQ1:} Is the Data Source Mapping activity perceived as useful for ideating ML-enabled systems? \textbf{RQ2:} Is the Feature-to-Data Source Mapping activity perceived as useful for ideating ML-enabled systems? \textbf{RQ3:} Is the ML Mapping activity perceived as useful for ideating ML-enabled systems? \textbf{RQ4:} Is the Define-ML approach perceived useful for ideating ML-enabled systems?

\subsection{Validation Context}

A three-hour session was conducted with a company from the Brazilian energy sector, integrated into their internal ideation training program. Eleven practitioners participated (IT, data analysis, and business strategy roles), where researchers presented the particularities of ideating ML-enabled systems and the Define-ML approach and its activities (one hour) and facilitated the practitioners in applying each of the three new activities (30 minutes for each) in practice on a toy problem. The toy problem concerned an intelligent banking system for loan approval. It is noteworthy that, as suggested by the \textit{Technology Transfer Model}~\cite{gorschek2006model}, the static validation did intentionally use a simulated context.

\subsection{Data Collection and Analysis Procedures}

After the session, the participants completed a printed questionnaire about their familiarity with ideation workshops, the perceived utility of the new activities and of Define-ML (using a five-point likert scale and open-text for feedback), and improvement suggestions.

\subsection{Results}

\begin{figure}[ht]
    \centering
    \includegraphics[width=0.5\textwidth]{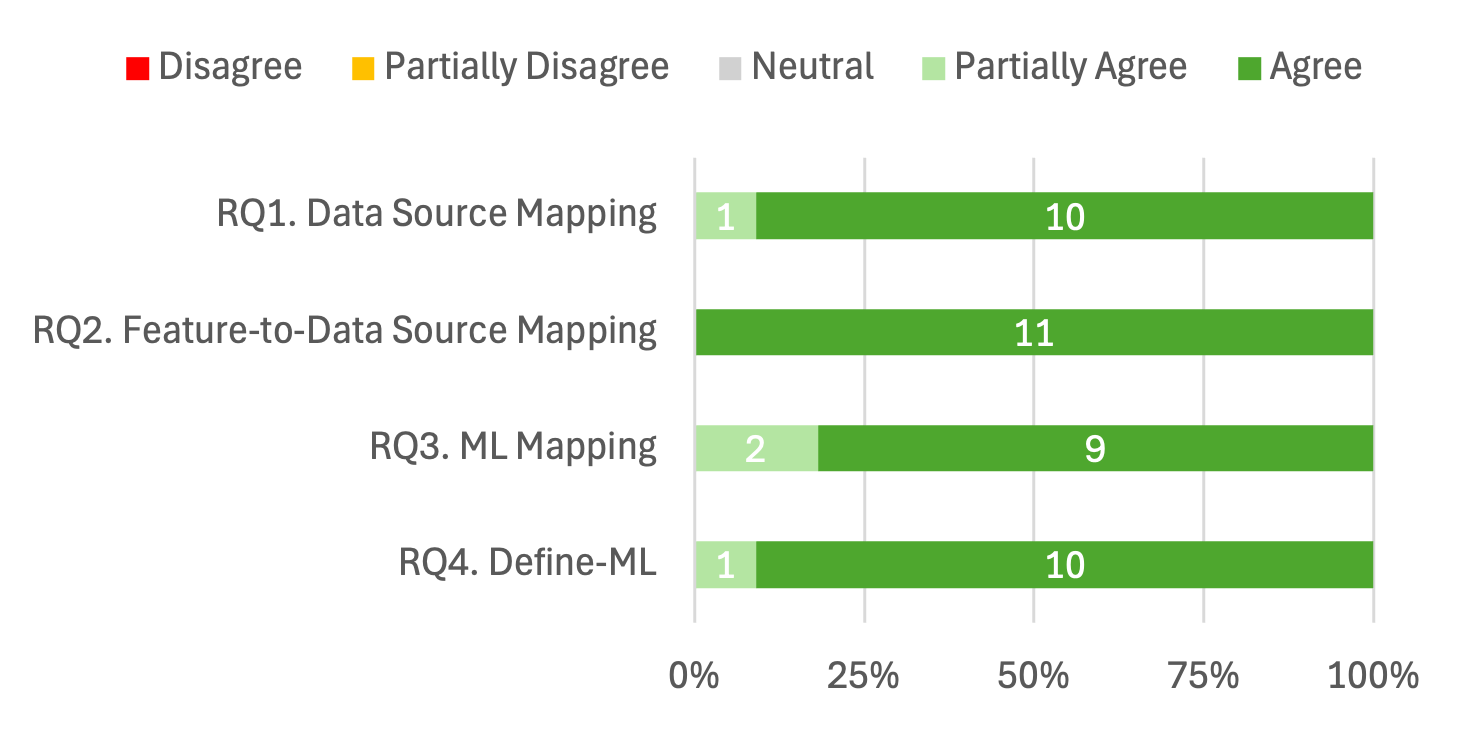}
    \caption{Agreement frequencies on usefulness}
    \label{fig:static_validation_results}
\end{figure}

The participants demonstrated varied exposure to ideation methods, with 2 participants (18\%) being highly experienced (three or more prior workshops), 6 (54\%) moderately experienced (one to two workshops), and 3 (27\%) first-time participants. All instruments and data are available in our open science repository. The quantitative analysis of the responses is presented in Figure~\ref{fig:static_validation_results}.

\subsubsection{\textbf{RQ1: Data Source Mapping Perceived Usefulness}} The majority of respondents (10 out of 11) agreed that the activity of mapping data sources is valuable for delineating ML-enabled system products. One participant partially agreed, indicating a minor divergence in opinion.

The feedback on the activity of mapping data sources was overwhelmingly positive. Participants emphasized that it is essential for understanding data architecture and that it plays a crucial role in enhancing the scope of solutions. The activity was noted as critical for comprehending data formats, maturity, and feasibility, which are key factors in ensuring the success of ML-enabled systems. Additionally, it was highlighted for promoting valuable discussions about data quality and its alignment with project expectations.


\subsubsection{\textbf{RQ2: Feature-to-Data Source Mapping Perceived Usefulness}} All 11 respondents unanimously agreed on the usefulness of mapping features to the data sources. This highlights strong support for the activity's importance in the process.

The feedback on mapping features with data highlighted its significant role in facilitating a clear understanding of functional expectations based on data. Participants noted that this activity helps align functionality with data sources, which not only reduces the likelihood of scope changes but also ensures that deliverables are realistic and well-defined.

\subsubsection{\textbf{RQ3: ML Mapping Perceived Usefulness}}
Nine respondents agreed that classifying data and identifying ML techniques is a useful activity. However, two participants partially agreed, reflecting some concerns regarding its applicability.

The feedback on classifying data and ML techniques emphasized the importance of supporting technical solution analysis and ensuring the feasibility of ML-intensive features. Participants noted that this activity enables a deeper understanding of data types and provides clarity on how to effectively treat and classify them. However, some challenges were identified, as certain respondents found the discussion overly technical and less relevant to business stakeholders. Indeed, a suggestion was made to simplify the data classification process, making it more accessible and aligned with business needs.

\subsubsection{\textbf{RQ4: Define-ML Perceived Usefulness}}
Ten respondents expressed agreement on the Define-ML framework’s usefulness for ideating ML-enabled system products. One participant partially agreed, indicating a minor divergence.

The feedback on the Define-ML framework highlighted its effectiveness. Participants appreciated the structured approach, which aids in aligning functionalities, data, and expectations to achieve cohesive outcomes.

%% file: 6_dynamic_validation.tex
\section{Dynamic Validation}\label{sec:dynamic_validation}

We report the dynamic validation as a case study following the guidelines suggested by Runeson \textit{et al.}~\cite{runeson2012case}.

\subsection{Goal and Research Questions}
We followed the GQM goal definition template~\cite{basili1988tame} to define the research goal as follows: ``\textit{Analyze} Define-ML \textit{for the purpose of} characterization \textit{with respect to} perceived usefulness of its activities and acceptance of the approach \textit{from the point of view of} practitioners \textit{in the context of} a real-world ML-enabled system product ideation."

Based on this goal, we defined the following RQs: \textbf{RQ1:} Is the Data Source Mapping activity perceived as useful for ideating ML-enabled systems? \textbf{RQ2:} Is the Feature-to-Data Source Mapping activity perceived as useful for ideating ML-enabled systems? \textbf{RQ3:} Is the ML Mapping activity perceived as useful for ideating ML-enabled systems? \textbf{RQ4:} How well is Define-ML accepted for the ideation of ML-enabled systems?

\subsection{Validation Context}

A three-day in-person Define-ML ideation workshop was scheduled and took place at the ExACTa PUC-Rio lab. The briefing provided to the ExACTa team set the context for the ideation session, explaining that the partner company, a large multinational energy drink manufacturer, sought to ``enhance their demand prediction capabilities."

The workshop attendees included a diverse group of professionals. The workshop was facilitated by the last author with direct support from the first author. Five additional specialists from the ExACTa lab provided technical ML expertise. Additionally, eleven professionals from the partner company participated, bringing domain knowledge and practical insights into the company’s business needs and operational challenges.

\begin{figure}[ht]
    \centering
    \includegraphics[width=0.45\textwidth]{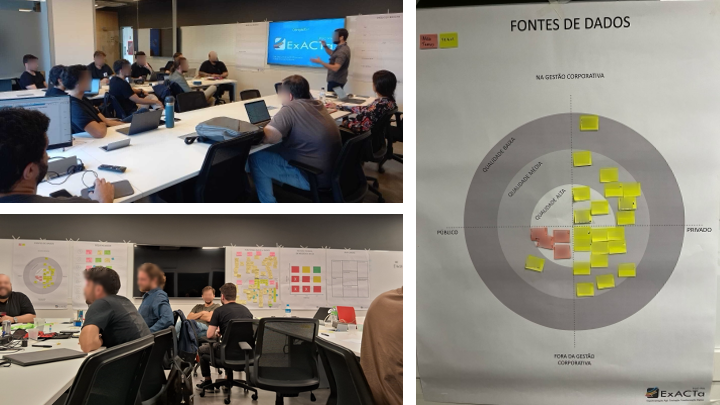}
    \caption{Workshop sessions and the printed Define-ML boards being filled}
    \label{fig:photos}
\end{figure}

Throughout the sessions, technical experts from ExACTa and professionals from the partner company worked together in co-creation (Figure~\ref{fig:photos}). The activities were supported by printed templates of the Define-ML boards, and the participants collaborated by filling post-its and attaching them to the boards. The second author manually copied the physical boards to a digital board in Miro.

\subsection{Data Collection and Analysis Procedures}

Participants were invited to complete a questionnaire via Google Forms, designed based on the Technology Acceptance Model (TAM)~\cite{davis1989perceived}. The questionnaire was developed to evaluate the research questions by capturing both quantitative and qualitative insights regarding participants' experiences and perceptions of the proposed ideation workshop framework.

To ensure ethical standards, the questionnaire provided information on the study's ethical considerations, specifically clarifying that all responses would remain anonymous and the identities of the participants would not be disclosed. This measure aimed to foster a sense of security and openness, encouraging honest and unbiased responses.

The questions concerned the participants' professional backgrounds and ideation workshop experience, TAM-based perceptions (utility, ease of use, intention to use) on Define-ML, and specific questions on the perceived usefulness of each new activity. Quantitative data from TAM-based measures (utility, ease of use, intention to use) were analyzed using descriptive statistics to identify participant acceptance trends. Qualitative responses underwent thematic analysis by the first author, with themes validated by the last author to ensure reliability.

\subsection{Results}

\begin{figure}[ht]
    \centering
    \includegraphics[width=0.5\textwidth]{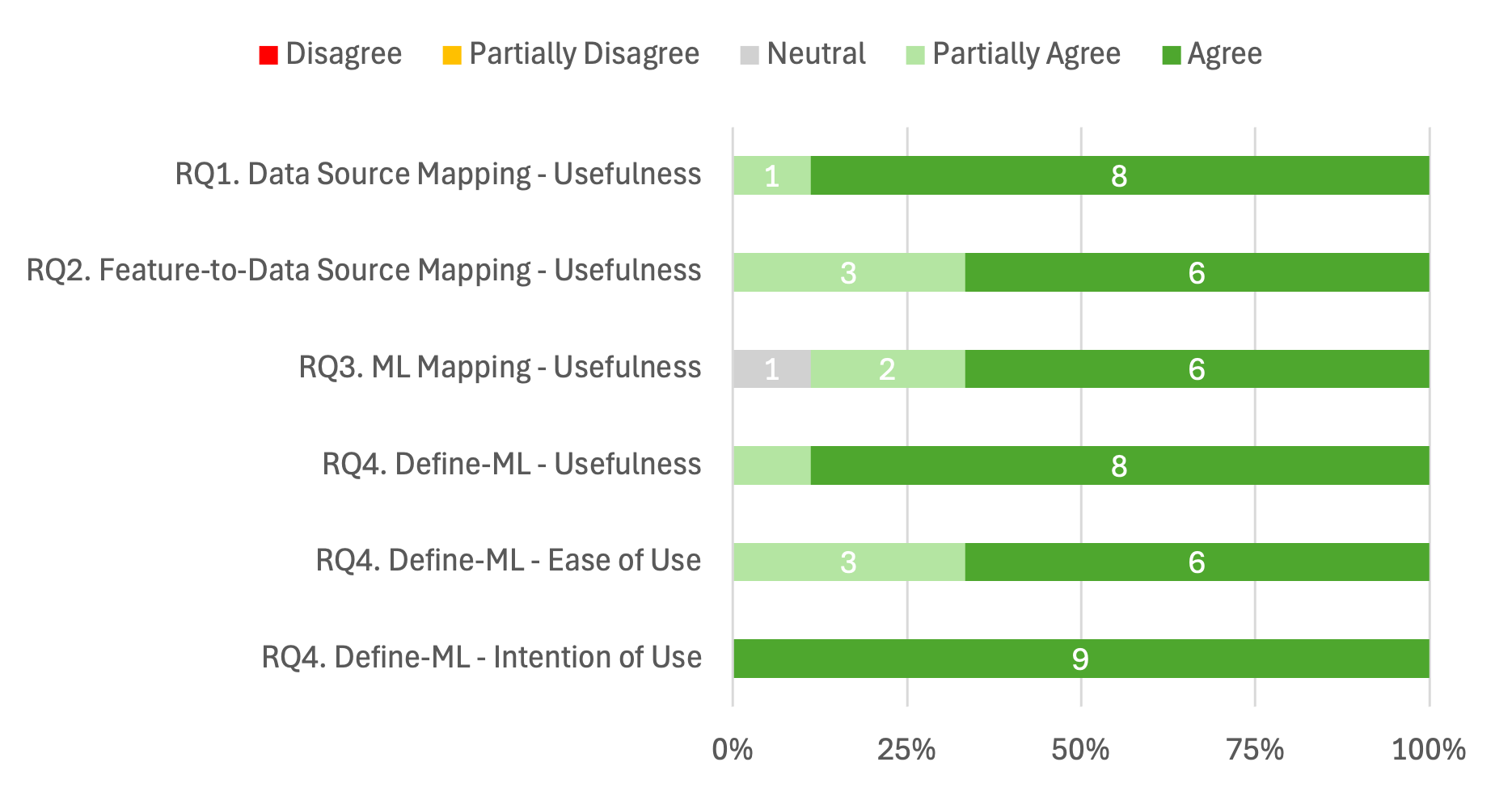}
    \caption{Frequencies of the responses}
    \label{fig:dynamic_validation_results}
\end{figure}

Filling out the questionnaire was not mandatory and only nine of the eleven business partner participants provided answers. Experience levels with ideation workshops varied significantly: six respondents (67\%) had participated in three or more previous workshops (e.g., Design Thinking, Lean Inception), while two (22\%) reported one prior experience, and one (11\%) had no previous exposure. This distribution provided valuable perspectives from both ideation-experienced practitioners and first-time participants. The quantitative analysis of the responses is shown in Figure~\ref{fig:dynamic_validation_results}.

\subsubsection{\textbf{RQ1: Data Source Mapping Activity}}
Quantitative analysis revealed that 8 out of 9 participants (89\%) agreed on the usefulness of the Data Source Mapping activity, with one participant expressing partial agreement. The themes that emerged from the qualitative analysis follow.

\textbf{Enhanced Data Understanding.} Participants valued the activity's ability to clarify data complexity. One noted: \textit{``Understanding all data sources and their qualities provided excellent clarity about current complexity.''} Another highlighted how it revealed \textit{``the volume of data processed by the company and the origin/utility of each source.''}

\textbf{Data Alignment.} The visual mapping enabled teams to design unified data strategies: \textit{``We easily created a blueprint for aligning data into a single source.''} Another participant appreciated the \textit{``visual artifact to group data and identify gaps to achieve product objectives.''}


As an improvement opportunity, one participant suggested avoiding data quality assessment at this stage: \textit{``Useful to show data complexity, but quality mapping felt inconsistent across stakeholders. Perhaps focus only on source identification without quality ratings.''}. This indicates potential to streamline the activity while preserving its core value.

In practice, the activity successfully reduced ambiguity about data availability, fostered cross-functional alignment. The strong agreement confirms its utility as an important element of ML ideation.

\subsubsection{\textbf{RQ2: Feature-to-Data Source Mapping Activity}}
The Feature-to-Data Source Mapping activity also received strong endorsement from participants, with 6 out of 9 practitioners (67\%) fully agreeing on its usefulness and 3 (33\%) partially agreeing. The analysis revealed two key themes:

\textbf{MVP Scope Definition.} Participants recognized the activity's role in clarifying deliverables, with one noting: \textit{``Necessary to understand the MVP's immediate benefit and the final product's value.''} Another emphasized how it \textit{``prevents creating features unsupported by available data.''}

\textbf{Data Impact Visualization.} The mapping helped understanding data-product relationships: \textit{``Useful to see how each data source impacts the product.''} Others valued how it \textit{``clarifies which data supports each feature for MVP prioritization.''}


Participants who partially agreed mentioned two considerations. One regarding technical familiarity: \textit{``Our team understands features well but needs more clarity on implementation paths and which data sources would be used.''} and other regarding the short time: \textit{``Because of the short ideation time... the task got a bit confusing.''}

Hence, while \textit{Feature-to-Data Source Mapping} was effective and recognized as useful to support precise scope definition and understand the impact of data, participants suggest allocating more time to maximize its effectiveness.

\subsubsection{\textbf{RQ3: ML Mapping Activity}}

The ML Mapping activity evaluation revealed mixed but generally positive perceptions among participants, with six out of nine practitioners (67\%) agreeing on its usefulness, 2 (22\%) partially agreeing, and 1 neutral (11\%). The analysis revealed three key themes:

\textbf{Technical Requirements Clarification.} Participants valued the activity's role in scoping ML needs: \textit{``Very useful to understand the ML process requirements for the final product.''} Another noted it \textit{``ensures chosen techniques align with specific product feature needs.''}

\textbf{Application Point Identification.} The mapping helped teams locate optimal ML integration points: \textit{``Useful to identify where ML model application would be valuable, considering data source limitations.''}

\textbf{Stakeholder Alignment.} The activity bridged technical-business gaps: \textit{``An opportunity for customer participation in typically technical discussions.''}; \textit{``Useful to align knowledge and expectations across the team.''} 

The participant who provided a neutral response justified: \textit{``This was the artifact that confused me most.''}. Another participant also recognized the complexity: \textit{``A quick workshop about ML techniques could happen before the activity.''}

Overall, we consider that the activity successfully introduced ML considerations into the ideation, but recognize that it requires careful facilitation to maximize its effectiveness across diverse stakeholder groups.

\subsubsection{\textbf{RQ4: Define-ML Acceptance}}

The evaluation of Define-ML's acceptance revealed strong agreement across all three TAM dimensions:

\textbf{Perceived Usefulness.} 
Eight out of nine participants (89\%) agreed that the approach effectively supports ML-enabled systems product ideation. They highlighted benefits such as structured collaboration—``it helped us jointly design the product and its core benefits; the in-person format was crucial”—as well as value acceleration by structuring and prioritizing deliverables, and cross-functional alignment by unifying diverse needs into data-driven solutions.

\textbf{Ease of Use.} Six participants (67\%) found the approach easy to use with proper facilitation, while three partially agreed. Participants noted that facilitators played a key role—\textit{``we couldn't have done it independently''}—and acknowledged initial complexity for first-timers, which eased over time. Others emphasized that \textit{``effective facilitation is needed to ensure productive contributions.''}

\textbf{Intention to Use.} All nine participants (100\%) expressed willingness to adopt Define-ML. They appreciated how it helps transform ideas into products and accelerates project outcomes. One participant noted, \textit{``A really pleasant experience seeing ideas become concepts then products.''} Another one mentioned that it \textit{``brings together both sides interested in product development, improving knowledge exchange.''}

%% file: 7_discussion.tex
\section{Discussion and Threats to Validity}\label{sec:discussion}

\subsection{Discussion}
The evaluation results indicate that Define-ML helps to bridge the gap between traditional product ideation and the requirements of ML-enabled systems. The aproach's three core innovations—\textit{Data Source Mapping}, \textit{Feature-to-Data Source Mapping}, and \textit{ML Mapping}—address critical challenges identified in prior work, particularly the alignment of technical capabilities with business objectives and the management of data dependencies in early-stage ideation.

Our findings support existing research on the importance of data-awareness in ML projects (\textit{e.g.}, \cite{nahar2023meta}\cite{alves2023status}) while introducing a structured approach to product-oriented ideation. The Data Source Mapping activity proved particularly valuable in making data visible and actionable early in the process, reducing the risk of later rework. Similarly, the Feature-to-Data Source Mapping gave practitioners a vision of what data sources need to be available to develop specific features. Finally, the ML Mapping activity successfully translated abstract ML concepts into practical product features.

Participants identified skilled facilitation (mentioned by 5 participants) as a critical element for successful adoption. They also praised the method’s collaborative approach, which fostered inclusivity, alignment, and stakeholder engagement, aiding decision-making and consensus-building. Overall, results indicate acceptance of Define-ML, with a unanimous intent to adopt the framework, suggesting high potential for industry adoption.

Compared to existing solutions, Define-ML advances beyond toolkit approaches~\cite{jansen2023mix} by embedding ML considerations within a full product ideation workflow. While feature catalogs~\cite{yildirim2023creating} can provide valuable inspiration, our framework offers systematic translation of features into implementable solutions through its integrated mapping activities.


%% file: 8_threats.tex
\subsection{Threats to Validity}\label{sec:threats}

Following Runeson \textit{et al.}~\cite{runeson2012case}, we examine potential threats in four validity categories.

\textbf{Construct validity:} Our approach combined quantitative surveys based on the TAM constructs with qualitative feedback. While TAM and qualitative feedback basically capture perceptions, we also triangulated these with the actual ideation outcomes (\textit{i.e.}, the results of the conducted activities). However, future work could benefit from longitudinal tracking of Define-ML adoption and impact.

\textbf{Internal validity:} Both validation settings included participants with varied levels of ML and ideation experience, which may have influenced responses. In addition, all workshops were facilitated by the researchers, which could introduce facilitator bias. To mitigate this, we collected anonymous feedback and included open-ended questions.

\textbf{External validity:} We validated Define-ML with partners from two industry domains using a toy problem and a real-world case. The dynamic validation reflects a single application context and company culture. Broader replications across additional domains, team compositions, and project types are necessary to establish external validity more robustly.

\textbf{Reliability:} Qualitative analysis was conducted by the first author and reviewed by the last author, with consensus used to resolve discrepancies. Furthermore, we strengthened reliability through methodological triangulation (\textit{e.g.}, comparing questionnaire data with actual workshop artifacts).



%% file: 9_conclusion.tex
\section{Conclusion}\label{sec:conclusion}

Define-ML provides a structured framework for addressing the unique challenges of ideating ML-enabled systems. By extending Lean Inception with targeted activities—\textit{Data Source Mapping, Feature-to-Data Source Mapping}, and \textit{ML Mapping}—the approach helps teams ground ideation in real data constraints, clarify ML feasibility, and align proposed features with business objectives.

Through static and dynamic validations with industry partners, Define-ML indicated strong perceived usefulness, ease of use when properly facilitated, and a clear practitioners' intent to adopt. The results suggest that Define-ML can help to effectively bridge gaps left by traditional ideation techniques. In contrast to existing toolkits or guidelines that focus on isolated aspects of ML design, Define-ML offers an end-to-end, workshop-based ideation approach that integrates data and model considerations into the collaborative sessions. These characteristics position Define-ML as a valuable contribution toward more actionable and feasible ML product definitions.

As future work, we aim to expand Define-ML’s scope to accommodate emerging AI paradigms, such as Generative AI and intelligent agents.